\documentclass[twocolumn,aps,pra,superscriptaddress,showpacs]{revtex4-1}
\usepackage{epsfig}
\usepackage{natbib}
\usepackage{color} 
\usepackage[utf8]{inputenc}
\usepackage[T1]{fontenc}
\usepackage{booktabs}
\usepackage{siunitx}

\usepackage{journalabbreviations}   % Abbreviations for journals used in *.bib file  

\usepackage{dcolumn}
\newcolumntype{d}[1]{D{.}{.}{#1}}
\usepackage{multirow}
\usepackage{rotating}
\usepackage{float}
\usepackage{amssymb}
\usepackage{enumitem} 
\usepackage{amsthm}
\usepackage{amsmath}

\newcommand{\blap}[1]{%
	\smash[c]{\begin{tabular}[c]{@{}c@{}}#1\end{tabular}}}

\definecolor{darkgreen}{rgb}{0.0, 0.6, 0.0}

\begin{document}

%\preprint{}

\title{New beyond-Voigt line-shape profile recommended for the HITRAN database}



\author{P. Wcis{\l}o}
\email[]{piotr.wcislo@umk.pl}
\affiliation{Institute of Physics, Faculty of Physics, Astronomy and Informatics, Nicolaus Copernicus University in Toru{\'{n}}, Grudzi{\k{a}}dzka 5, 87-100 Toru{\'{n}}, Poland}

\author{N. Stolarczyk}
\affiliation{Institute of Physics, Faculty of Physics, Astronomy and Informatics, Nicolaus Copernicus University in Toru{\'{n}}, Grudzi{\k{a}}dzka 5, 87-100 Toru{\'{n}}, Poland}

\author{M. Słowiński}
\affiliation{Institute of Physics, Faculty of Physics, Astronomy and Informatics, Nicolaus Copernicus University in Toru{\'{n}}, Grudzi{\k{a}}dzka 5, 87-100 Toru{\'{n}}, Poland}

\author{H. Jóźwiak}
\affiliation{Institute of Physics, Faculty of Physics, Astronomy and Informatics, Nicolaus Copernicus University in Toru{\'{n}}, Grudzi{\k{a}}dzka 5, 87-100 Toru{\'{n}}, Poland}

\author{D. Lisak}
\affiliation{Institute of Physics, Faculty of Physics, Astronomy and Informatics, Nicolaus Copernicus University in Toru{\'{n}}, Grudzi{\k{a}}dzka 5, 87-100 Toru{\'{n}}, Poland}

\author{R. Ciuryło}
\affiliation{Institute of Physics, Faculty of Physics, Astronomy and Informatics, Nicolaus Copernicus University in Toru{\'{n}}, Grudzi{\k{a}}dzka 5, 87-100 Toru{\'{n}}, Poland}

\author{A. Cygan}
\affiliation{Institute of Physics, Faculty of Physics, Astronomy and Informatics, Nicolaus Copernicus University in Toru{\'{n}}, Grudzi{\k{a}}dzka 5, 87-100 Toru{\'{n}}, Poland}

\author{F. Schreier}
\affiliation{Remote Sensing Technology Institute, German Aerospace Center (DLR), Wessling, Germany}

\author{C.D. Boone}
\affiliation{Department of Chemistry, University of Waterloo, 200 University Avenue West, Ontario N2L 3G1, Canada}

\author{A. Castrillo}
\affiliation{Department of Mathematics and Physics, Università degli Studi della Campania “Luigi Vanvitelli”, 81100, Caserta, Italy}

\author{L. Gianfrani}
\affiliation{Department of Mathematics and Physics, Università degli Studi della Campania “Luigi Vanvitelli”, 81100, Caserta, Italy}

\author{Y. Tan}
\affiliation{Hefei National Research Center for Physical Sciences at the Microscale, University of Science and Technology of China, Hefei, China}

\author{S-M. Hu}
\affiliation{Hefei National Research Center for Physical Sciences at the Microscale, University of Science and Technology of China, Hefei, China}

\author{E. Adkins}
\affiliation{Chemical Sciences Division, National Institute of Standards and Technology, Gaithersburg, MD, USA}

\author{J.T. Hodges}
\affiliation{Chemical Sciences Division, National Institute of Standards and Technology, Gaithersburg, MD, USA}

\author{H. Tran}
\affiliation{Laboratoire de Météorologie Dynamique/IPSL, CNRS, Ecole polytechnique, Institut Polytechnique de Paris, Sorbonne Université, Ecole Normale Supérieure, Université PSL, F-91120 Palaiseau, France }

\author{H.N. Ngo}
\affiliation{Faculty of Physics, Hanoi National University of Education, 136 Xuan Thuy Street, Cau Giay District, Hanoi, Vietnam}
\affiliation{Institute of Natural Sciences, Hanoi National University of Education, 136 Xuan Thuy Street, Cau Giay District, Hanoi, Vietnam}

\author{J.-M. Hartmann}
\affiliation{Laboratoire de Météorologie Dynamique/IPSL, CNRS, Ecole polytechnique, Institut Polytechnique de Paris, Sorbonne Université, Ecole Normale Supérieure, Université PSL,
F-91120 Palaiseau, France }

\author{S. Beguier}
\affiliation{University of Grenoble Alpes, CNRS, LIPhy, Grenoble F-38000, France}

\author{A. Campargue}
\affiliation{University of Grenoble Alpes, CNRS, LIPhy, Grenoble F-38000, France}

\author{R.J. Hargreaves}
\affiliation{Harvard-Smithsonian Center for Astrophysics, Atomic and Molecular Physics Division, Cambridge, MA 02138, USA}

\author{L.S. Rothman}
\affiliation{Harvard-Smithsonian Center for Astrophysics, Atomic and Molecular Physics Division, Cambridge, MA 02138, USA}

\author{I.E. Gordon}
\affiliation{Harvard-Smithsonian Center for Astrophysics, Atomic and Molecular Physics Division, Cambridge, MA 02138, USA}


\date{\today}

\begin{abstract}

Parameters associated with the collisional perturbation of spectral lines are essential for modeling the absorption of electromagnetic radiation in gas media.  The HITRAN molecular spectroscopic database provides these parameters, although originally they were associated only with the Voigt profile parameterization.  However, in the HITRAN2016 and HITRAN2020 editions, Voigt, speed-dependent Voigt and Hartmann-Tran (HT) profiles have been incorporated, thanks to the new relational structure of the database. The HT profile was introduced in HITRAN in 2016 as a recommended profile for the most accurate spectral interpretations and modeling. It was parameterized with a four-temperature-range temperature dependence. Since then, however, some features of the HT profile have been revealed that are problematic from a practical perspective. These are: the singular behavior of the temperature dependencies of the velocity-changing parameters when the shift parameter crosses zero and the difficulty in evaluating the former for mixtures.  In this article, we summarize efforts to eliminate the above-mentioned problems that led us to recommend using the quadratic speed-dependent hard-collision (qSDHC) profile with double-power-law (DPL) temperature dependencies. We refer to this profile as a modified Hartmann-Tran (mHT) profile. The computational cost of evaluating it is the same as for the HT profile. We give a detailed description of the mHT profile (also including line mixing) and discuss the representation of its parameters, together with their DPL temperature parametrization adopted in the HITRAN database. We discuss an efficient algorithm for evaluating this profile and provide corresponding computer codes in several programming languages: Fortran, Python, MATLAB, Wolfram Mathematica, and LabVIEW. We also discuss the associated update of the HITRAN Application Programming Interface (HAPI).

\end{abstract}

%Although the parameters of any line-shape model can be stored in the HITRAN database, in this paper we recommend to use the quadratic speed-dependent hard-collision (qSDHC) profile which compromises a simple structure of the model and accessibility
%of efficient computer algorithm with ability to accurately represent the shapes of molecular lines (involving both speed dependence of broadening and shift, and velocity-changing collisions). The qSDHC profile is similar to the Hartmann-Tran profile; qSDHC profile adopts modifications that makes it more suitable for molecular spectra representation. The goal of this paper is to provide a detailed description of the qSDHC profile and discuss the representaion of its parameters adopted in the HITRAN database. We discuss the efficient algorithm for evaluating the qSDHC profile and provide a corresponding Fortran computer code; tez ze HAPI. 

%\pacs{33.70.Jg, 42.50.Xa, 42.62.Fi}
%42.50.Xa	Optical tests of quantum theory
%42.62.Fi   Laser spectroscopy
%33.70.Jg	Line and band widths, shapes, and shifts
%34.20.Gj	Intermolecular and atom-molecule potentials and forces

%\keywords{}

\maketitle

\section{Introduction}

A model of the collision-induced shape of spectral lines is needed for the accurate reproduction of observed spectra from individual line parameters provided in spectroscopic databases. On the one hand, the Voigt profile often cannot reproduce experimental line shapes to better than percent accuracy at typical atmospheric pressures and temperatures (see e.g. \cite{Duggan1997,Pine1999,DeVisia2011,Long2011,Ngo2013,Lisak2015a}). Additionally, the requirement for accuracy, e.g. in remote sensing of atmospheric composition, has already reached the sub-percent level \cite{Miller2007}. In particular, measurements of greenhouse gas concentrations require highly accurate reference data, which can be achieved only with sophisticated beyond-Voigt models of line shapes, e.g.~\cite{Delahaye2016,Long2021}. Many other fields related to gas metrology also require accurate beyond-Voigt line-shape models, e.g., hygrometry \cite{Bielska2012,Hashiguchi2022}, remote pressure and temperature measurements \cite{Smith1989,Wehr2003,Castrillo2019}, and even testing \textit{ab initio} quantum chemical calculations \cite{Slowinski2020,Bielska2022} and quantum electrodynamics for molecules \cite{Zaborowski2020}.

To address the problem raised above, a variety of sophisticated line-shape models were developed over the last decades. They aimed at taking into account two main beyond-Voigt physical effects: speed dependence of the pressure-induced width and shift modeled with the hypergeometric \cite{Berman1972,Ward1974} or quadratic approximation \cite{Rohart1994} and the changes of the radiator velocity induced by collisions modeled with the hard- \cite{Nelkin1964,Rautian1966,Robert1993,Lance1997,Pine1999,DeVizia2014} or soft-collision \cite{Galatry1961,Ciurylo1997,Ciurylo2001SDGP} approximation. Combinations of hard- and soft-collision models were also discussed, extending the approaches of Rautian and Sobelman \cite{Rautian1966,Lance1998,Lance1999,Ciurylo1998,Ciurylo2001JQSRT,Shapiro2001} as well as Keilson and Storer \cite{Keilson1952,Robert1998,Robert2000,Shapiro2001,Bonamy2004}  (see review in \cite{Hartmann2021book}). 
A more realistic rigid-sphere model \cite{Lindenfeld1980,Liao1980} of velocity-changing collisions takes into account the mass ratio of colliding molecules and incorporates the relative importance of the speed- and direction-changing collisions \cite{Ciurylo2002BB,Ciurylo2002H2,Wcislo2014,Wcislo2015PRA2}. This model is used in the billiard-ball (BB) and speed-dependent billiard-ball (SDBB) profiles \cite{Blackmore1987,Ciurylo2002BB}. A computationally efficient algorithm to calculate billiard-ball profiles \cite{Wcislo2013PRA} enables applications to a wide range of gas pressures, down to the Doppler limit. The advantage of the rigid sphere approach over simpler models is particularly pronounced in the case of molecular hydrogen lines \cite{Wcislo2015PRA2, Wcislo2018, Zaborowski2020,Slowinski2020}. Finally, velocity-changing collisions were modeled with a repulsive inverse-power potential \cite{Blackmore1987} in the case of speed-dependent spectral line shapes \cite{Wcislo2013JQSRT}.
Among these line-shape models, the partially-correlated speed-dependent hard-collision profile \cite{Pine1999} with quadratic speed-dependent pressure-induced width and shift \cite{Rohart1994} turned out to have a simple algebraic form which is a combination of Voigt profiles \cite{Ngo2013,Tran2013}. This property makes it very efficient from a computational perspective. Moreover, this profile can mimic more elaborated models with good accuracy \cite{Wcislo2017JPCS}. 

The partially-correlated quadratic speed-dependent hard-collision profile (pCqSDHCP) was suggested to be used in future databases in 2013 \cite{Ngo2013}, considering its universality, low computational cost \cite{Tran2013} and ability to reproduce experimental spectra to below 1 permille accuracy. This profile was later-on recommended by the IUPAC (International Union of Pure and Applied Chemistry) water-vapor task group as the new beyond-Voigt line shape standard for spectroscopic databases \cite{Tennyson2014PAC}, and it was named the Hartmann-Tran (HT) profile. The authors of the original work subsequently diagnosed an inconsistency in the sign convention assumed in the original formulation \cite{Ngo2014,Tran2014} and conclusively clarified the issue in a note \cite{Hartmann2019}. In 2016, the HT profile was implemented in the HITRAN database \cite{Wcislo2016JQSRT,Kochanov2016,Gordon2017}. To improve the accuracy in the representation of temperature dependencies of the line-shape parameters, the full temperature range (from $0$ to $1000$~K) was split into four intervals (4TR representation) and within each of them, a separate set of parameters associated with a single power law in temperature was provided \cite{Wcislo2016JQSRT}.

Since its original formulation \cite{Ngo2013}, the HT profile has been widely used for accurate spectra simulations and analysis (e.g.~\cite{Loos2014,Adkins2021}) which allowed the community to identify some limitations related to the parametrization of the velocity-changing collision part. 
In the HT profile, the complex rate, $\nu_{opt}$, that scales the hard-collision velocity-changing operator is parameterized in the following way
\begin{equation}
\nu_{opt}=\nu_{VC}-\eta(\Gamma_0+i\Delta_0)-\eta(\Gamma_2+i\Delta_2)(v^2/v_m^2-3/2),
    \label{eq:HTPnuOpt}
\end{equation}
where $\nu_{VC}$ is the frequency of
velocity-changing collisions and $\eta$ the correlation parameter. $\Gamma_0$ and $\Delta_0$ are the speed-averaged pressure broadening and shift, and $\Gamma_2$ and $\Delta_2$ are the corresponding speed dependence parameters. The actual and most probable speeds of an active molecule are denoted as $v$ and $v_m$, respectively. Although the definition and detailed discussion of the mHT profile are given in Sec.~\ref{sec:theProfile}, here we give an expression defining $\nu_{opt}$ for the mHT profile,
\begin{equation}
\nu_{opt}=\nu_{opt}^r+i\nu_{opt}^i,
    \label{eq:qSDHCnuOpt}
\end{equation}
to highlight a key difference between the two profiles. $\nu_{opt}^r$ and $\nu_{opt}^i$ are the real and imaginary parts of the Dicke narrowing parameter \cite{Stolarczyk2020}. The much more complicated way of defining $\nu_{opt}$ in the HT profile (which also involves the $\Gamma_0$ and $\Delta_0$, and $\Gamma_2$ and $\Delta_2$ parameters) causes some problems for spectral analysis and for storing these parameters in the database.

\begin{figure}
    \centering
    \includegraphics[width=\linewidth]{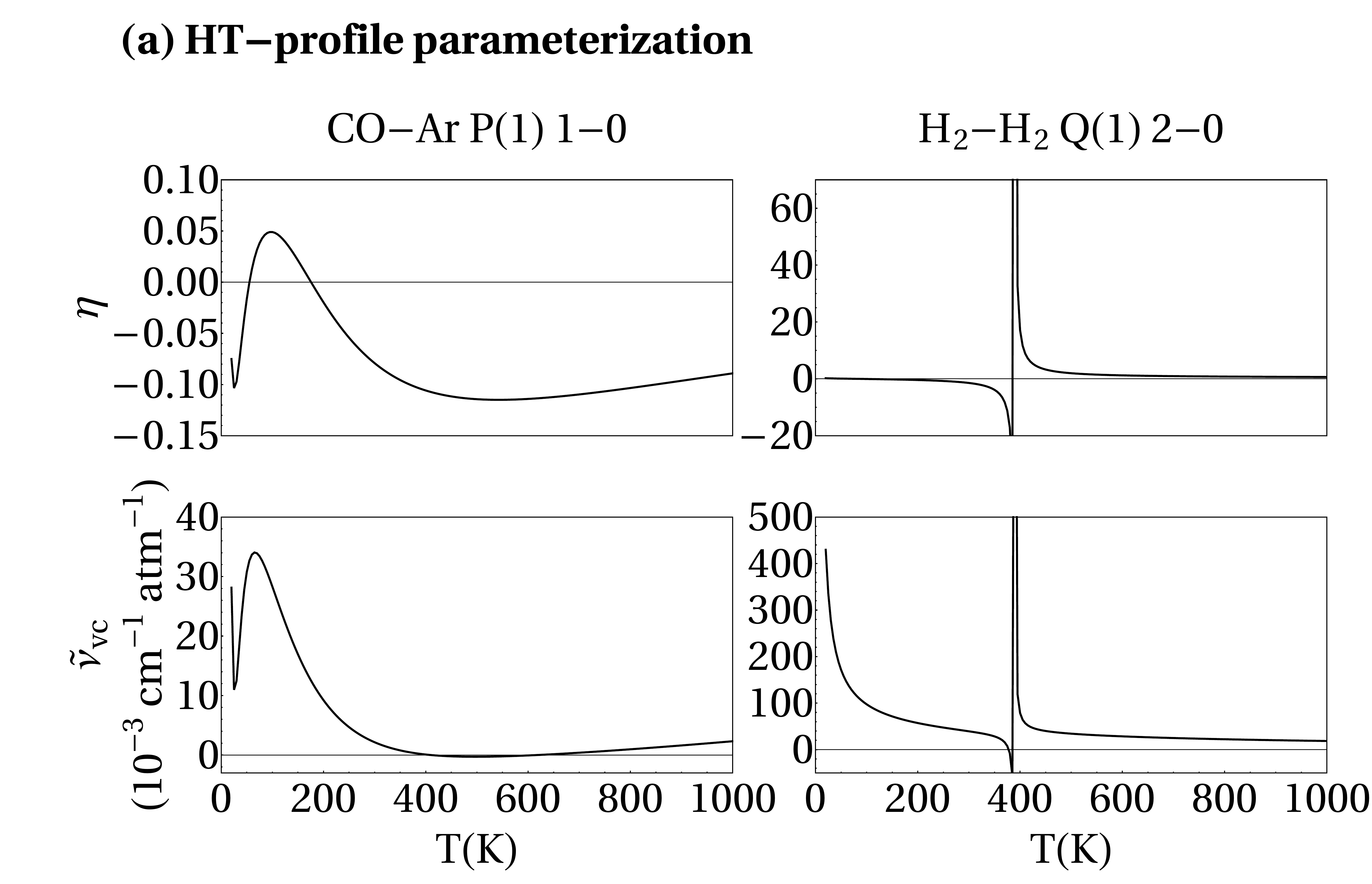}
    \includegraphics[width=\linewidth]{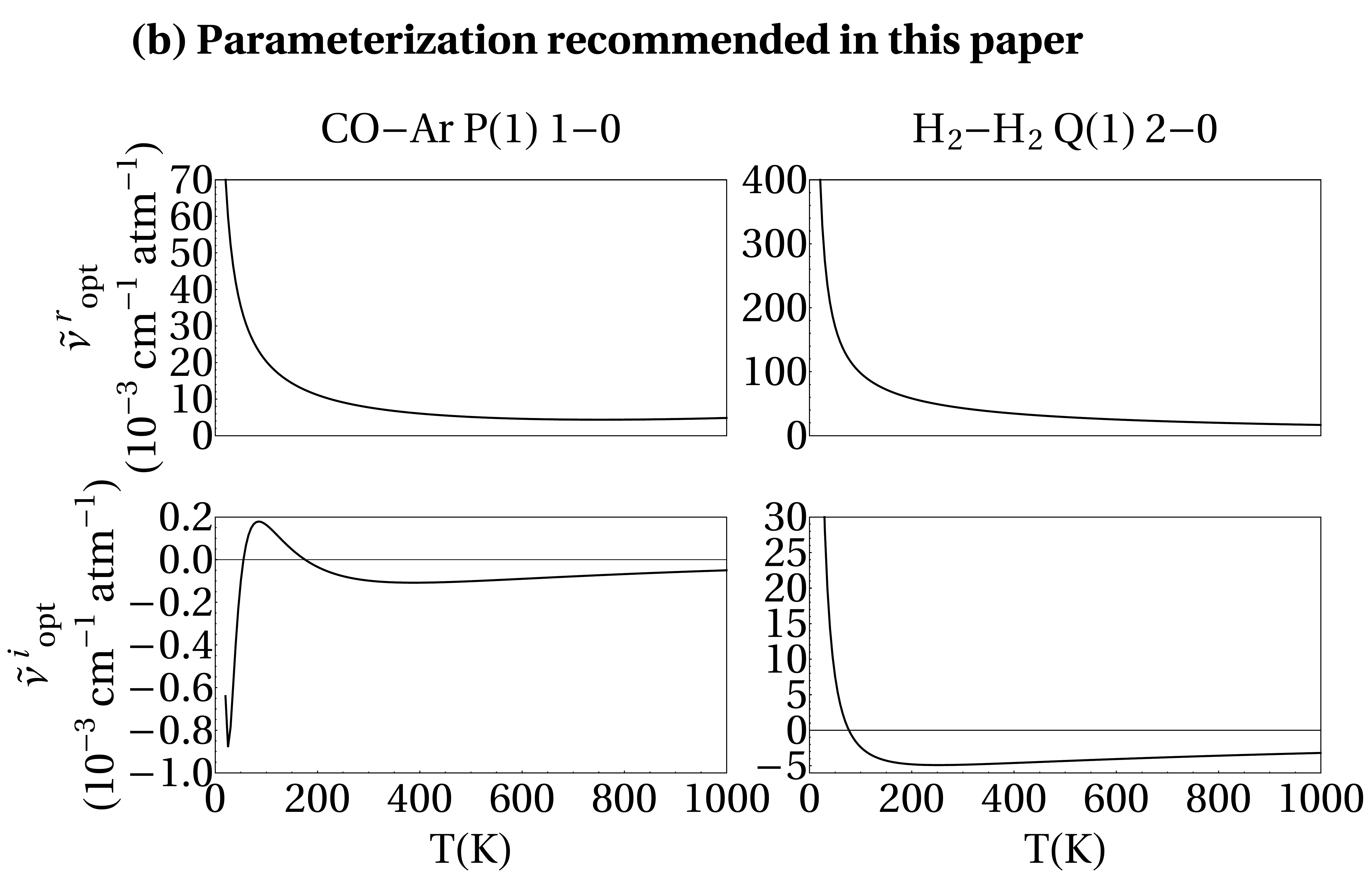}
    \caption{Examples of temperature dependencies of the velocity-changing line-shape parameters. (a) HT-profile parametrization: the frequency of velocity-changing collisions, $\tilde{\nu}_{VC}$ (tilde denotes the pressure-independent version of $\nu_{VC}$, i.e., $\nu_{VC}=p~\tilde{\nu}_{VC}$), and the correlation parameter $\eta$. (b) Parameterization recommended in this paper based on a direct decomposition of the complex Dicke parameter into its real, $\tilde{\nu}_{opt}^r$, and imaginary, $\tilde{\nu}_{opt}^i$, parts.}
    \label{fig:tempDepend}
\end{figure}

A direct comparison of Eqs.~(\ref{eq:HTPnuOpt}) and (\ref{eq:qSDHCnuOpt}) cannot be made because Eq.~(\ref{eq:HTPnuOpt}) introduces an additional dependence of $\nu_{opt}$ on the speed of the absorbing molecule (the problem of the speed dependence of $\nu_{opt}$ is discussed in the end of this section). Nonetheless, to explain the origin of the singular behavior of the $\nu_{VC}$ and $\eta$ parameters, one can ignore the last term (the speed dependence of $\nu_{opt}$) in Eq.~(\ref{eq:HTPnuOpt}) and compare the two equations, which allows the expression of $\nu_{VC}$ and $\eta$ in terms of $\nu_{opt}^r$ and $\nu_{opt}^i$ \cite{Stolarczyk2020}:
\begin{equation}
\nu_{VC}=\nu_{opt}^r-\nu_{opt}^i\frac{\Gamma_0}{\Delta_0},
    \label{eq:trans1}
\end{equation}
\begin{equation}
\eta=-\frac{\nu_{opt}^i}{\Delta_0}.
    \label{eq:trans2}
\end{equation}
It directly results from these equations that the HT profile parameters, $\nu_{VC}$ and $\eta$, exhibit singular behaviors when $\Delta_0$ crosses zero. We show in Fig.~\ref{fig:tempDepend}~(a) an example of a system, the Q(1) 2-0 line in self-perturbed H$_2$, that exhibits such behavior. The black curves in the right plots in panel (b) show the temperature dependencies of the real and imaginary parts of the Dicke narrowing parameter ($\nu_{opt}^r$ and $\nu_{opt}^i$). Both parameters exhibit a nice and smooth behavior versus temperature and are well approximated with the double-power-law (DPL) representation. However, when we transform these temperature dependencies into the HT profile parametrization (using Eqs.~(\ref{eq:trans1}) and (\ref{eq:trans2})), then the singularities appear at the point where the shift parameter, $\Delta_0$, crosses zero, see Fig.~\ref{fig:tempDepend}. It is very problematic to represent such temperature dependencies in a database. It should be emphasized that the singular behavior of $\nu_{VC}$ and $\eta$ does not originate from the properties of the complex Dicke narrowing parameter but from the parametrization of the HT profile (see Eq.~(\ref{eq:HTPnuOpt})). 

The temperature dependencies of the HT profile parameters $\nu_{VC}$ and $\eta$ can also be problematic in some other cases in which $\Delta_0$ does not cross zero. Examples of it are shown in the left panels in Fig.~\ref{fig:tempDepend} (a) and (b) for the case of the P(1) 1-0 line in Ar-perturbed CO. The temperature dependence of the $\nu_{opt}^r$ parameter is well reproduced by a DPL, whereas the $\nu_{opt}^i$ parameter exhibits a strange behavior at low temperatures that makes it difficult to be well approximated with a DPL. The magnitude of $\nu_{opt}^i$ is much smaller than $\nu_{opt}^r$ and errors in representing the temperature dependence of $\nu_{opt}^i$ have almost negligible influence on the line shape. However, when we use the HT profile parametrization, much of the strange behavior of $\nu_{opt}^i$ also propagates to $\nu_{VC}$ which makes it difficult to be represented with the DPL (see the lower left panel in Fig.~\ref{fig:tempDepend} (a)), resulting in a huge impact on the line shape. 

The second problematic feature of the parametrization of the HT profile is the nontrivial way of evaluating the $\nu_{VC}$ and $\eta$ for mixtures. For the mHT profile, all the line-shape parameters for a mixture are calculated as a simple weighted sum of the contributions from the different perturbers, see Sec.~\ref{sec:gasMixture}. However, for the HT profile, the complicated definition of the complex Dicke parameter (Eq.~(\ref{eq:HTPnuOpt})) implies a nontrivial way in which the $\nu_{VC}$ and $\eta$ parameters are calculated for mixtures (see Appendix~\ref{sec:app1} for derivation): 
\begin{equation}
    \eta=\frac{\sum\limits_i\eta_i(\gamma_{2,i}+i\delta_{2,i})p_i}{\sum\limits_i(\gamma_{2,i}+i\delta_{2,i})p_i},
    \label{eq:etaSum}
\end{equation}
\begin{equation}
    \nu_{VC}=\sum\limits_i\left[\tilde{\nu}_{VC,i}+(\gamma_{0,i}+i\delta_{0,i})(\eta-\eta_i)\right]p_i,
    \label{eq:nuVCSum}
\end{equation}
where $p_i$ is the partial pressure of the $i$-th constituent. Equations~(\ref{eq:etaSum}) and (\ref{eq:nuVCSum}) are exact and do not involve any approximation. They are, however, problematic from a practical perspective. The summation rule for mixtures is different and more complicated than for all the other line-shape parameters. This would make the database less intuitive to use and might potentially lead to some confusion and mistakes.

Besides addressing the two above-mentioned problematic features stemming from the parametrization of the HT profile, there are some other advantages of the mHT profile parametrization related to its physical meaning. First, the explicit decomposition of $\nu_{opt}$ into its real, $\nu_{opt}^r$, and imaginary, $\nu_{opt}^i$, parts offers direct access to their line-shape interpretation, see the detailed discussion in Appendix~A of Ref.~\cite{Stolarczyk2020}. Second, within the HT profile the speed dependence of the complex Dicke narrowing parameter is enforced by the speed dependencies of the broadening and shifting parameters, which is not physical, see Appendix~B of Ref.~\cite{Stolarczyk2020}.

The HT profile parameterization implemented in the HITRAN 2016 database \cite{Wcislo2016JQSRT,Kochanov2016, Gordon2017} was based on temperature dependencies of the line-shape parameters split into four temperature ranges (4TR representation). However, it was demonstrated in Ref.~\cite{Stolarczyk2020} that the double-power-law (DPL, defined in~\cite{Gamache2018} for $\gamma_0$ and $\delta_0$ and generalized for other line-shape parameters in~\cite{Stolarczyk2020}) gives a better overall approximation of the temperature dependencies than does the 4TR. The DPL also requires fewer parameters and its structure is much simpler and more self-consistent~\cite{Stolarczyk2020}. 

In this article, we summarize the efforts made to solve the problems outlined in the above paragraphs, which leads us to recommend using the quadratic speed-dependent hard-collision (qSDHC) profile with DPL temperature dependencies. To keep consistency with previous HITRAN notation~\cite{Gordon2017}, we refer to this profile as a modified Hartmann-Tran (mHT) profile. In Section~\ref{sec:theProfile}, we give a detailed description of the mHT profile (Sec.~\ref{sec:isolated}), including discussions of the $\beta$ correction function (Sec.~\ref{sec:betaCorr}), line mixing (Sec.~\ref{sec:lineMixing}), and gas mixtures (Sec.~\ref{sec:gasMixture}; the details on gas mixtures are provided in Appendix~\ref{sec:app1}). In Section~\ref{sec:numericalEvaluation}, we present an efficient algorithm for evaluating the mHT profile, and provide the corresponding computer codes in several programming languages: Fortran, Python, MATLAB, Mathematica, and LabVIEW. These codes are available in the Supplementary Materials to this work and at the GitHub platform \cite{github_2024}. In Section~\ref{sec:HITRANparametersAndHAPI}, we discuss the representation of the mHT profile parameters together with their DPL temperature parametrization which will be included in the HITRAN2024 official release~\cite{Gordon2025}, and we discuss the corresponding updates of the HITRAN Application Programming Interface (HAPI). Section~\ref{sec:conclusion} summarizes this work.

\section{New beyond-Voigt line-shape profile recommended for the HITRAN database}
\label{sec:theProfile}

In addition to the basic collisional effects that are included in the Voigt profile (pressure shift and broadening and the Doppler broadening of a line), the \mbox{mHT} profile includes the effects of speed dependence of the broadening, shift, and of velocity-changing collisions. The speed dependencies are approximated with quadratic functions and velocity-changing collisions are described using the hard-collision model (within this model, the velocity is completely thermalized after each collision). 

In Section \ref{sec:isolated}, we define all the line-shape parameters of the mHT profile for the case of isolated lines (in the format adopted in HITRAN2024~\cite{Gordon2025}) and we give explicit formulas to evaluate the profile. The hard-collision approximation is sufficient for most of the molecular systems that are present in HITRAN. However, for light molecules, for which Dicke narrowing is pronounced, the hard-collision approximation can be insufficient. To handle this complication we use a simple correction factor, the $\beta$-correction, which does not require any additional parameter, and which is given by a simple analytical function, see Sec.~\ref{sec:betaCorr}. In Section~\ref{sec:lineMixing}, we discuss how the mHT profile is treated in HITRAN when line-mixing is present. In Section~\ref{sec:gasMixture}, we discuss the case of gas mixtures.\\

\subsection{The isolated line case}
\label{sec:isolated}

We denote the mHT profile as $I_{mHT}(\nu)$. The profile is normalized, which means that
\begin{equation}
    \label{eq:profileNormalization}
    \int I_{mHT}(\nu)d\nu=1.
\end{equation}
The mHT profile involves six collisional line-shape parameters:
\begin{equation}
    \label{eq:parameters}
    \Gamma_0,\Delta_0,\Gamma_2,\Delta_2,\nu_{opt}^r,\nu_{opt}^i,
\end{equation}
which are provided in the HITRAN database~\cite{Gordon2025} in the pressure-normalized form:
\begin{align}
    \label{eq:parametersA}
    \begin{split}
    \gamma_0 & =\Gamma_0/p,~ \delta_0=\Delta_0/p\\
    \gamma_2 & =\Gamma_2/p,~ \delta_2=\Delta_2/p\\
    \tilde{\nu}_{opt}^r & =\nu_{opt}^r/p,~ \tilde{\nu}_{opt}^i=\nu_{opt}^i/p,
    \end{split}
\end{align}
and are given in the units of cm$^{-1}$atm$^{-1}$. Additionally, the mHT profile involves the Doppler-broadening parameter
\begin{equation}
    \Gamma_D=\sqrt{\ln 2}~\nu_D,
\end{equation}
where $\nu_D=\nu_0 v_m/c$, $c$ is the speed of light, and $\nu_0$ is the unperturbed frequency of the transition. $\Gamma_0$ and $\Gamma_D$ are the half width at half maximum (HWHM) of the Lorentz and Doppler broadening components, respectively.

The mHT profile, $I_{mHT}(\nu)$, can be calculated as
\begin{equation}
    \label{eq:qSDHC}
    I_{mHT}(\nu)=Re[\tilde{I}_{mHT}(\nu)],
\end{equation}
where $\tilde{I}_{mHT}(\nu)$ is a complex mHT function that is given by
\begin{equation}
    \label{eq:complexqSDHC}
    \tilde{I}_{mHT}(\nu)=\frac{\tilde{I}^{*}_{qSDV}(\nu)}{1-(\nu_{opt}^r+i\nu_{opt}^i)\pi\tilde{I}^{*}_{qSDV}(\nu)},
\end{equation}
where $\tilde{I}_{qSDV}(\nu)$ is a  complex quadratic speed-dependent Voigt profile that is given by
\begin{align}
    \label{eq:complexqSDV}
    \begin{split}
    \tilde{I}_{qSDV}(\nu)&=\\
    \frac{1}{\pi}\int  d^3\vec{v}&f_m(\vec{v})\frac{1}{\Gamma(v)+i\Delta(v)-i(\nu-\nu_0-\nu_D v_z/v_m)},
    \end{split}
\end{align}
where $v_z$ is the $z$-component of an active molecule velocity, $\vec{v}$, $f_m(\vec{v})=(\sqrt{\pi}v_m)^{-3}e^{-(\vec{v}/v_m)^2}$ is the Maxwellian velocity distribution of $\vec{v}$ and $\Gamma(v)+i\Delta(v)$ is a quadratic function
\begin{equation}
\label{eq:quadraticFunkction}
    \Gamma(v)+i\Delta(v)=\Gamma_0+i\Delta_0+(\Gamma_2+i\Delta_2)(v^2/v^2_m-3/2).
\end{equation}
The asterisk in Eq.~(\ref{eq:complexqSDHC}) denotes that $\Gamma(v)+i\Delta(v)$ in Eq.~(\ref{eq:complexqSDV}) is replaced with $(\Gamma(v)+\nu_{opt}^r)+i(\Delta(v)+\nu_{opt}^i)$. The expression for $\tilde{I}^{*}_{qSDV}(\nu)$ can be written in an explicit form 
\begin{widetext}
\begin{align}
    \label{eq:complexqSDV2}
    \begin{split}
    \tilde{I}^{*}_{qSDV}(\nu)=\frac{1}{\pi}\int  d^3\vec{v}f_m(\vec{v})
    \frac{1}{\Gamma_0+i\Delta_0+(\Gamma_2+i\Delta_2)(v^2/v^2_m-3/2)+\nu_{opt}^r+i\nu_{opt}^i-i(\nu-\nu_0-\nu_D v_z/v_m)}.
    \end{split}
\end{align}
\end{widetext}
Equations (\ref{eq:qSDHC}), (\ref{eq:complexqSDHC}) and (\ref{eq:complexqSDV2}) give a direct recipe for evaluating the mHT profile.
 
It should be noted that the mHT profile can be obtained from the original HT profile \cite{Ngo2013,Tennyson2014PAC} by setting the correlation parameter, $\eta$, to zero and generalizing the frequency of velocity-changing collisions, $\nu_{VC}$, to a complex number. For most of the previous measurements in which the HT profile was used, the $\eta$ parameter was either very small or even set to zero. Therefore, such data obtained from the HT profile can be straightforwardly used in the mHT parametrization.

\subsection{The $\beta$ correction function}
\label{sec:betaCorr}

The hard-collision model of velocity-changing collisions, which is used in the mHT profile, aptly handles the velocity-changing line-shape effects (such as the Dicke narrowing) for the majority of molecular species. However, in the cases with a prominent degree of Dicke narrowing (e.g., molecular hydrogen transitions) the hard-collision model does not reproduce line shapes at the required accuracy level. To overcome this problem, a simple analytical correction (the $\beta$ correction function) was introduced \cite{Wcislo2016JQSRT,Konefal2020} that allows one to mimic (with the mHT profile) the behavior of the more physical billiard-ball model \cite{Blackmore1987,Ciurylo2002BB} and, hence, to considerably improve the accuracy of the mHT profile in this (prominent Dicke narrowing) regime at negligible numerical cost. The correction is made by replacing $\nu_{opt}^{r}$ with $\beta_\alpha(\chi) \nu_{opt}^{r}$, where $\alpha$ is the perturber-to-absorber mass ratio and $\chi=\nu_{opt}^{r}/\Gamma_D$ (see Ref.~\cite{Konefal2020} for details). It should be emphasized that the $\beta$ correction does not require any additional transition-specific parameter (it depends only on the perturber-to-absorber mass ratio  $\alpha$). 

To date, the $\beta$ correction has been widely tested in numerous experiments and simulations~\cite{Bielska2022, Castrillo2021, Lamperti2023a, Wcislo2021, Gotti2021, Zaborowski2020, Mondelain2023, Fleurbaey2021, Wojtewicz2020, Adkins2022, Bielska2021, Domyslawska2022, Kowzan2020, Le2024, Odintsova2020, Lamperti2023b, TranDD2021, Jozwiak2024, Lamperti2023c} often demonstrating that it outperforms the non-corrected profile. In Section~\ref{sec:mHTComputerRoutines}, we discuss practical details regarding how the $\beta$ correction is implemented in our mHT routines.

\subsection{Line mixing}
\label{sec:lineMixing}

Accurate modeling of molecular spectra at atmospheric conditions often requires including the line mixing effect \cite{Hartmann2021book,Tran2010,Benner2016}. Line mixing arises from collisional coupling \cite{Hartmann2021book,Baranger1958,Kolb1958,Fano1963,BenReuven1966b} of molecular states corresponding to different optical transitions and is especially prominent in cases of closely spaced molecular lines \cite{Pine1997CH4,Hartmann1999,Vitcu2004,Flaud2006,Domyslawska2022} and high-pressure spectra \cite{Ozanne1997,Sheldon1998HD}, but also in the troughs between lines
(as it is seen in atmospheric spectra of CO$_2$ and O$_2$ used by for OCO, see e.g. Refs.~\cite{Tran2008,Hartmann2009}. 

Molecular spectra affected by speed-dependent effects \cite{Berman1972}, Dicke narrowing \cite{Dicke1953}, and line mixing \cite{Baranger1958,Pine1997sum} can be described within the framework of the theory developed by Smith {et al.} \cite{Smith1971i,Smith1971ii}. Within the hard collision model  \cite{Nelkin1964,Rautian1966} the molecular spectrum takes the form of matrix generalizations \cite{Ciurylo2000} of the speed-dependent hard-collision profiles. In the general case, such a spectrum cannot be represented by an ordinary sum of separately parameterized individual lines with their parameters varying linearly with pressure. Instead, the band of overlapping lines should be evaluated as a whole. 

This theory and its simplification were successfully used in the analysis of experimental data \cite{Pine2000,Pine2003,Pine2004,Pine2019}. Strong line mixing leads to a nonlinear pressure dependence of line-shape parameters and variation of the line intensity. On the other hand when line mixing is weak in the first order approximation \cite{Rosenkranz1975}, assuming speed-independent off-diagonal elements describing the collisional coupling between lines, the general expression of Ref. \cite{Ciurylo2000} can be reduced to a sum of asymmetric speed-dependent hard-collision profiles \cite{Ciurylo1998} with line shape parameters that vary linearly with pressure \cite{Pine2000,Ciurylo2000,Pine2003} and with unperturbed line intensities.
The role of speed dependence of line mixing in data fitting was also recently discussed in Ref.~\cite{Boulet2021}. On the other hand, Thibault et al. \cite{Thibault2022} showed in their calculations that the speed dependence of off-diagonal elements of the relaxation matrix can be weak. This finding was also supported by results from \cite{Serov2021} where speed dependence of line mixing asymmetry was not experimentally accessible.
The asymmetric speed-dependent hard-collision profile \cite{Ciurylo1998} is a linear combination of the speed-dependent hard collision profile and its dispersive component.

%It should be noted, however, that the dispersive component of spectral line shape can also originate from the finite time of molecular collision which leads to the breakdown of an impact approximation. Spectral line shape \cite{Ciurylo1997,Ciurylo1998,Ciurylo2001JQSRT} including collision-time asymmetry \cite{Anderson1955,Szudy1977,Royer1978,Julienne1986,Szudy1996,Ciurylo1997asym} is mathematically the same as the one including the first-order line mixing. Simultaneous treatment of the collision-time asymmetry and line mixing, in the general case, was given in Ref.~\cite{Ciurylo2001LM} and the simple additivity of both effects was derived in the weak line mixing case. Experimental studies and theoretical estimations of collision-time effects (also called non-impact effects) in molecular systems indicate that at atmospheric conditions the intensities of the spectral lines can be affected by this effect at the permille level \cite{Marteau1984HF,Boulet2004HCl,Reed2023PRL}. However, in some specific systems, these non-impact effects can be even larger. 

Summing up, in the regime of weak line coupling, line mixing can be easily handled by the mHT profile (without any further numerical cost) by combining its real and imaginary parts
\begin{equation}
    \label{eq:qSDHCLM}
    I_{mHT}(\nu)=(1-iY)\tilde{I}_{mHT}(\nu),
\end{equation}
% ta konwecja w przypadku Lorentza daje profil:
% [(1-X)+Y(w-w0-D)]/[G^2+(w-w0-D)^2]
where $Y$ is the first-order Rozenkranz line-mixing parameter \cite{Rosenkranz1975}, and its pressure-normalized counterpart is denoted as $y=Y/p$. It should be noted that there are two possible sign conventions related to the line-mixing parameter~\cite{Hartmann2019}. Although both yield correct results, the sign of the $Y$ contribution in Eq.~\eqref{eq:qSDHCLM} must be the same as the sign of the $i(\nu-\nu_0-\nu_D v_z/v_m)$ part of Eq.~\eqref{eq:complexqSDV2}. Suppose both signs are negative, as in the case of the present article. In that case, one can identify the real part of $I_{mHT}(\nu)$ as the absorption coefficient, and its imaginary part as the dispersion coefficient, but with a negative sign. In the other convention, with both positive signs, the absorption coefficient is unchanged and the dispersion coefficient is equal to the imaginary part of the $I_{mHT}(\nu)$ without the sign change.

%{\textcolor{red}{RC: Now we have opposite convention comparing to what was written before: In this work and in the attached mHT computer codes, we use the more straightforward approach with two signs being positive.}}

\begin{figure}[t]
    \centering
    \includegraphics[width=\linewidth]{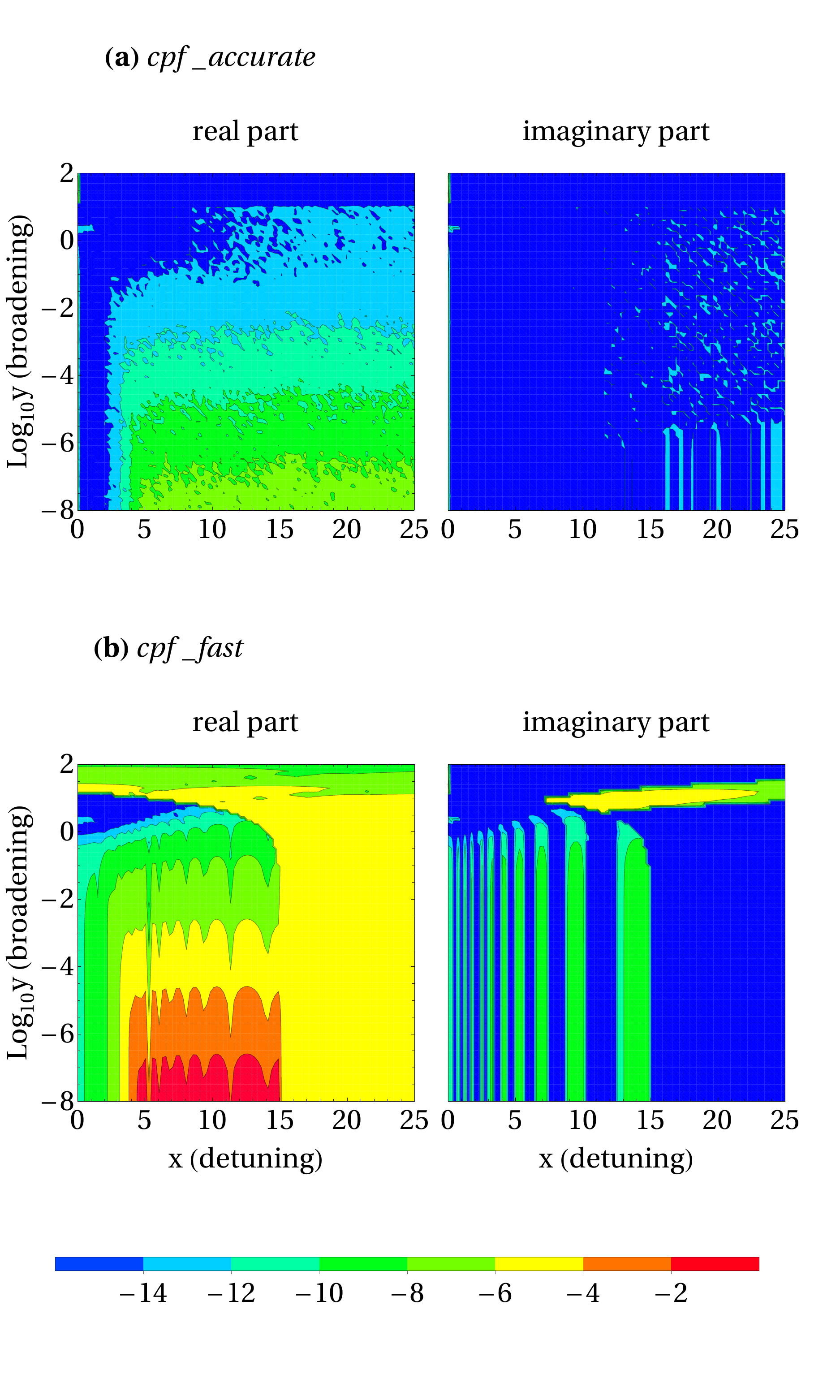}
    \caption{The relative accuracy of the complex probability function (CPF) routines (i.e., the routines that calculate the complex probability function, $w(z)$, where $z=x+i y$). In the Voigt-profile limit, the $x$ and $y$ parameters have the physical meaning of detuning from line center and Lorentzian broadening (both normalized to Doppler broadening, $\nu_D$), respectively. The color scale represents the relative accuracy of the routines expressed in orders of magnitude, e.g., -15 corresponds to a relative accuracy better than 10$^{-15}$.  The reference CPF for generating these plots was calculated based on direct integration of Eq.~\eqref{eq:CPF_definifion}.}
    \label{fig:accuracy}
\end{figure}

\subsection{Formulas for gas mixture}
\label{sec:gasMixture}

In the mHT profile, all the line-shape parameters for a mixture are calculated as a simple weighted sum of the contributions from different perturbers, i.e.:
\begin{align}
    \label{eq:parametersB}
    \begin{split}
    \Gamma_0 & =\sum\limits_j p_j\gamma_{0,j}, {~}{~}{~}\Delta_0 = \sum\limits_j p_j \delta_{0,j},\\
    \Gamma_2 & =\sum\limits_j p_j\gamma_{2,j}, {~}{~}{~}\Delta_2 = \sum\limits_j p_j \delta_{2,j},\\
    \nu_{opt}^r & =\sum\limits_j p_j\tilde{\nu}_{opt,j}^r,{~}{~}{~} \nu_{opt}^i =\sum\limits_j p_j\tilde{\nu}_{opt,j}^i,\\
    Y &=\sum\limits_j p_jy_{j},
    \end{split}
\end{align}
where $p_j$ is the partial pressure of the $i$-th species in the mixture. Note that in the case of the HT profile, due to the unfortunate parametrization of the complex Dicke narrowing parameter, the expressions for these parameters for mixtures are much more complicated, see Eqs.~(\ref{eq:etaSum}) and (\ref{eq:nuVCSum}), and the derivation in Appendix~\ref{sec:app1}.

\section{Numerical evaluation of the profile and fast computer routines for different programming languages}
\label{sec:numericalEvaluation}

\begin{table*}[]
    \centering
    \begin{tabular}{cSSSS}
         \multicolumn{5}{c}{CPU timing 11th Gen Intel Core i9-11900K @ 3.50 GHz}  \\
         \hline
         &\text{$\pm$25FWHM}&\text{$\pm$3FWHM}&\text{$\pm$25FWHM}&\text{$\pm$3FWHM}  \\
         &\multicolumn{2}{c}{\textit{cpf\_accurate}}&\multicolumn{2}{c}{\textit{cpf\_fast}} \\
         \cline{2-5}
         Python 3.12.3 &0.32&0.32&0.26&0.29  \\
         Fortran 18 &0.085&0.086&0.046&0.054 \\
         FORTRAN77 &0.087&0.087&0.032&0.051  \\
         MATLAB 9.13&1.5&1.5&0.6&1.0  \\    
         Mathematica 14&7.1&7.1&6.3&5.1  \\    
         LabVIEW 2025 Q1&0.24&0.24&0.11&0.16  \\
         \cline{2-5}
         &\multicolumn{2}{c}{\textit{weideman(64)}}&\multicolumn{2}{c}{\textit{hum1\_wei}} \\
         \cline{2-5}
         Python 3.12.3 (HAPI)&69.6&67.6&46.2&69.6  \\
    \end{tabular}
    \caption{A comparison of the single-thread CPU timings (in seconds) of the CPF evaluation with different programming languages. We tested the \textit{cpf\_accurate} and \textit{cpf\_fast} algorithms described in Sec.~\ref{sec:errorFunction}, and the \textit{hum1\_wei} and \textit{weideman(64)} functions from the hapi.py package~\cite{Kochanov2016}, as their corresponding HAPI references. The calculations were performed on the same 1000$\times$1000 grids, either focusing only on the line core ($\pm$3~full width at half maximum, FWHM) or also including the line wings ($\pm$25~FWHM). For the \textit{cpf\_accurate} routine the evaluation time is nearly the same for both ranges. For the \textit{cpf\_fast} routine  the evaluation time is slightly longer for the line core (in this range the simple Humlíček's algorithm cannot be used). All the routines presented in the table (except the HAPI library) are appended as Supplementary Material to this work and are available at the GitHub platform \cite{github_2024}.}
    \label{tab:timings}
\end{table*}

A key advantage of the mHT profile, from the perspective of its efficient numerical evaluation, is that it can be expressed as a quotient of two qSDV profiles, see Eq.(\ref{eq:complexqSDHC}). It is so, because (as demonstrated by~\citet{Boone2007}) the qSDV profile can be expressed as a simple sum of two complex probability functions (CPFs). Therefore, the problem of fast evaluation of the mHT profile reduces to having an efficient algorithm for evaluating the complex probability function.

\subsection{qSDV profile expressed as a sum of complex probability functions (CPFs)}
\label{sec:efficientAlgorithm}

It directly follows from the definition of a simple Voigt profile that it can be expressed in terms of the CPF. The problem gets complicated when the broadening and shift parameters depend on the speed of an active molecule. In general, to evaluate the profile value for a given light frequency, $\nu$, the integral in Eq.~(\ref{eq:complexqSDV}) has to be numerically calculated. It was, however, demonstrated in Ref.~\cite{Boone2007} that in the case of quadratic speed dependence (see Eq.~(\ref{eq:quadraticFunkction})) the integral can be expressed as a sum of two CPFs, $w(z)$, and Eq.~\eqref{eq:complexqSDV2} can be written as
\begin{equation}
\label{eq:Zform}
     \tilde{I}^{*}_{qSDV}(\nu)=\frac{1}{\pi \nu_D} \left(w\left(i Z_1\right)-w\left(i Z_2\right)\right).
\end{equation}
In this work, we mostly follow the notation from Ref.~\cite{Tran2013} (see also Ref.~\cite{Ngo2013}):
\begin{equation}
     Z_1=\sqrt{\frac{C_0-i(\nu-\nu_0)}{C_2} +\left(\frac{\nu_D}{2 C_2}\right)^2}-\frac{\nu_D}{2 C_2},
\end{equation}
\begin{equation}
     Z_2=\sqrt{\frac{C_0-i(\nu-\nu_0)}{C_2} +\left(\frac{\nu_D}{2 C_2}\right)^2}+\frac{\nu_D}{2 C_2},
\end{equation}
%{\color{red} RC: In Eqs. (19) and (20) "i" was repleaced by "-i" to keep used convetion.} 
where we follow the convention in which the pressure shift and its speed dependence enter the model with a positive sign~\cite{Tran2014,Hartmann2019}
\begin{equation}
     C_0=\Gamma_0+i\Delta_0+\nu_{opt}^r+i\nu_{opt}^i-\frac{3}{2}(\Gamma_2+i\Delta_2)
\end{equation}
and
\begin{equation}
     C_2=\Gamma_2+i\Delta_2.
\end{equation}

\subsection{Optimal algorithm for evaluating the CPF}
\label{sec:errorFunction}

The complex probability function (CPF) is defined as
\begin{equation}
     w(z)=e^{-z^2}\textrm{erfc}(-iz),
\end{equation}
where $z$ is a complex number, and $\textrm{erfc}(-iz)$ is the complementary error function, that is defined as
\begin{equation}
    \textrm{erfc}(-iz)=1+\frac{2 i}{\sqrt{\pi}}\int^z_0e^{-\tilde{z}^2}d\tilde{z}.
\end{equation}
For $\textrm{Im}(z)>0$ (which is the case for the line shape applications considered here, meaning the value of the pressure broadening is positive), the above integral over a complex argument can be written as an integral over a real argument~\cite{abramowitz1964handbook}.
\begin{equation}
    \textrm{erfc}(-iz)=\frac{i}{\pi}e^{z^2}\int^\infty_{-\infty}\frac{e^{-t^2}}{z-t}dt,
    \label{eq:CPF_definifion}
\end{equation}
which constitutes one of the simplest recipes for evaluating the CPF.

Due to the importance of the CPF in many fields, considerable efforts have been made over the past several decades to develop a fast and accurate algorithm for evaluating it. In 1975, a Taylor series expansion and continued fraction approach were reported~\cite{Drayson1975}. A few years later, Humlíček developed the rational approximation algorithm~\cite{Humlicek1978,Humlicek1982}. 
Note that the approaches from both Ref.~\cite{Drayson1975} and Refs.~\cite{Humlicek1978,Humlicek1982} divide the complex plane (spanned by the Voigt broadening and detuning parameters) into regions in which different methods of evaluating the CPF are used. The state of knowledge in the evaluation of the CPF (before 1993) was reported in Ref.~\cite{Thompson1993} addressing issues that include: analytical approximations, power series expansions, direct numerical integration, Fourier-transform-based methods, and those based on interpolations. Further tests, extensions and improvements of the Humlíček algorithm were published in Refs.~\cite{Schreier1992,Kuntz1996,Ryuten2004,Schreier2016, Schreier2018}. In 2007, the Lagrange polynomial was used for the region of small broadening and detuning~\cite{Letchworth2007}.
In 1994, Weideman introduced self-recurred rational expansions~\cite{Weideman1994} that were further used as part of other algorithms~\cite{Schreier2011,Schreier2021}.

The Fortran computer code for evaluating the HT profile reported in Ref.~\cite{Tran2013} is based on Humlíček's asymptotic approximation~\cite{Humlicek1982}. An algorithm by Schreier~\cite{Schreier2011}, which combines the Humlíček~\cite{Humlicek1982} approach and Weideman's~\cite{Weideman1994} 24-term rational approximation, was the default CPF used for the HT profile calculation in the HITRAN Application Programming Interface (HAPI)~\cite{Kochanov2016}.

\begin{center}
	\begin{table*}
		\caption{The parametrization of the DPL temperature dependence of the beyond-Voigt line-shape parameters.}			
		\small{
			\begin{tabular}{ c c c c c c c }
				\hline
				\multirow{6}{*}{\blap{Description of \\ the parameters}} & \multirow{6}{*}{\blap{ The notation of the\\ line-shape parameter\\.[cm$^{-1}$/atm]}} & \multicolumn{4}{c}{\multirow{4}{*}{\blap{DPL parametrization \\ of the temperature dependence\\of the line-shape parameters}}} & \multirow{6}{*}{\blap{The formulas illustrating\\how the parameters should be\\translated into the line-shape \\ parameters ($\rm T_{\rm ref}=$ 296~K) }}\\
				& & & & & \\
				& & & & & \\
				& & & & & \\
				\cline{3-6}
				& & Coef. 1 & Coef. 2 & Exp. 1 & Exp. 2 \\		 
				\hline
				\multirow{2}{*}{\blap{Pressure broadening}} & \multirow{2}{*}{\blap{$\gamma_0(\rm T)$}} & \multirow{2}{*}{\blap{ g$_0$}} & \multirow{2}{*}{\blap{ g$_0'$}} & \multirow{2}{*}{\blap{ n}} & \multirow{2}{*}{\blap{ n$'$}}&\multirow{2}{*}{\blap{$\gamma_0(\rm T)=$ g$_0({\rm T_{\rm ref}}/\rm T)^n$+g$_0'({\rm T_{\rm ref}}/\rm T)^{n'}$}}\\
				& & & & & \\ 
				\hline 
				\multirow{2}{*}{\blap{Pressure shift}} & \multirow{2}{*}{\blap{$\delta_0(\rm T)$}} & \multirow{2}{*}{\blap{ d$_0$}} & \multirow{2}{*}{\blap{ d$_0'$}} & \multirow{2}{*}{\blap{ m}} & \multirow{2}{*}{\blap{ m$'$}}&\multirow{2}{*}{\blap{$\delta_0(\rm T)=$ d$_0({\rm T_{\rm ref}}/\rm T)^m$+d$_0'({\rm T_{\rm ref}}/\rm T)^{m'}$}}\\
				& & & & & \\ 
				\hline 
				
				\multirow{2}{*}{\blap{Speed dependence of\\the pressure broadening}} & \multirow{2}{*}{\blap{$\gamma_2(\rm T)$}} & \multirow{2}{*}{\blap{ g$_2$}} & \multirow{2}{*}{\blap{ g$_2'$}} & \multirow{2}{*}{\blap{ j}} & \multirow{2}{*}{\blap{ j$'$}}&\multirow{2}{*}{\blap{$\gamma_2(\rm T)=$ g$_2({\rm T_{\rm ref}}/\rm T)^j$+g$_2'({\rm T_{\rm ref}}/\rm T)^{j'}$}}\\
				& & & & & \\ 
				\hline 
				\multirow{2}{*}{\blap{Speed dependence of\\the pressure shift}} & \multirow{2}{*}{\blap{$\delta_2(\rm T)$}} & \multirow{2}{*}{\blap{ d$_2$}} & \multirow{2}{*}{\blap{ d$_2'$}} & \multirow{2}{*}{\blap{ k}} & \multirow{2}{*}{\blap{ k$'$}}&\multirow{2}{*}{\blap{$\delta_2(\rm T)=$ d$_2({\rm T_{\rm ref}}/\rm T)^k$+d$_2'({\rm T_{\rm ref}}/\rm T)^{k'}$}}\\
				& & & & & \\ 
				\hline 
				
				\multirow{2}{*}{\blap{Real part\\ of the Dicke parameter}} & \multirow{2}{*}{\blap{$\widetilde{\nu}_{\rm opt}^{\rm r}(\rm T)$ }} & \multirow{2}{*}{\blap{ r}} & \multirow{2}{*}{\blap{ r$'$}} & \multirow{2}{*}{\blap{ p}} & \multirow{2}{*}{\blap{ p$'$}}&\multirow{2}{*}{\blap{$\widetilde{\nu}_{\rm opt}^{\rm r}(\rm T)=$ r$({\rm T_{\rm ref}}/\rm T)^p$+r$'({\rm T_{\rm ref}}/\rm T)^{p'}$}}\\
				& & & & & \\ 
				\hline 
				
				\multirow{2}{*}{\blap{Imaginary part\\ of the Dicke parameter}} & \multirow{2}{*}{\blap{$\widetilde{\nu}_{\rm opt}^{\rm i}(\rm T)$}} & \multirow{2}{*}{\blap{ i}} & \multirow{2}{*}{\blap{ i$'$}} & \multirow{2}{*}{\blap{ q}} & \multirow{2}{*}{\blap{ q$'$}}&\multirow{2}{*}{\blap{$\widetilde{\nu}_{\rm opt}^{\rm i}(\rm T)=$ i$({\rm T_{\rm ref}}/\rm T)^q$+i$'({\rm T_{\rm ref}}/\rm T)^{q'}$}}\\
				& & & & & \\ 
				\hline

    			\multirow{2}{*}{\blap{First-order Rosenkranz\\ line-mixing parameter}} & \multirow{2}{*}{\blap{$y(\rm T)$ }} & \multirow{2}{*}{\blap{ y$_0$}} & \multirow{2}{*}{\blap{ y$_0'$}} & \multirow{2}{*}{\blap{ e}} & \multirow{2}{*}{\blap{ e$'$}}&\multirow{2}{*}{\blap{$y(\rm T)=$ y$_0({\rm T_{\rm ref}}/\rm T)^e$+y$_0'({\rm T_{\rm ref}}/\rm T)^{e'}$}}\\
				& & & & & \\ 
				\hline 

			\end{tabular}	
		}
		
		\label{tab:DPL}	
	\end{table*}
\end{center}

\begin{figure*}
    \centering
    \includegraphics[width=\textwidth]{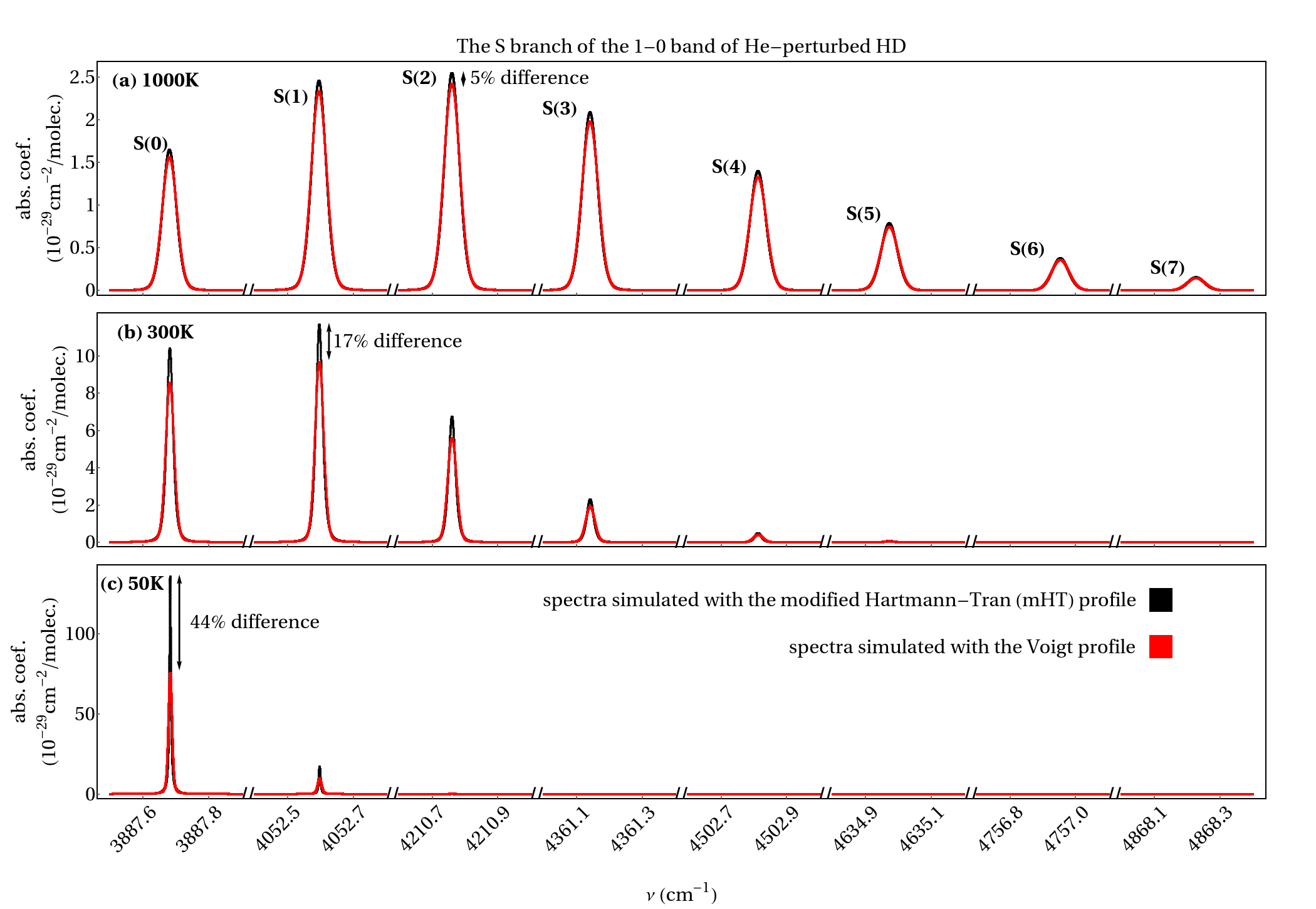}
    \caption{Example of spectra generated with the mHT profile (black curve) with the DPL temperature dependence parametrization for the case of the S branch of the 1-0 band in HD perturbed by He at three different temperatures at $p$~=~0.2~atm. The spectra were generated with HAPI (the  line-shape parameters were calculated in Ref.~\cite{Stankiewicz2020}). The same spectra generated with the Voigt profile (red curves) show that for some cases it is critical to properly handle in the database the beyond-Voigt line-shape effects (at low temperatures the profiles can differ by as much a factor of 2 for this system). 
    }
    \label{fig:Sbranch}
\end{figure*}

In this work, we reviewed the CPF routines known in the literature from the perspective of developing optimal CPF and mHT algorithms and, in particular, from the perspective of their applications in HAPI \cite{Kochanov2016}. We found that the 64-term Weideman rational approximation~\cite{Weideman1994} provides the best available compromise between accuracy and speed of the computer routine, and we use it as a starting point for developing our CPF routine dedicated for mHT calculations. We also consider Schreier's algorithm~\cite{Schreier2011} which is faster (but less accurate) to develop a fast version of our CPF routine.
\par
Both approaches require the evaluation of Weideman's rational approximation series table (obtained via fast Fourier transform, FFT), and several constants, which depend only on the number of approximation terms. For the mHT profile calculation, we fix the number of terms in Weideman's series, which allowed us to significantly accelerated our computational routines by pre-calculating the constant factors (in particular the FFT-based factors). We found that, for the mHT calculation, reducing the number of Weideman terms from 64 to 42 offers acceleration with a negligible loss in accuracy. Furthermore, in the 42-term case, because the initial five terms are small, we truncated them, while maintaining the same accuracy but further accelerating the calculations. Therefore, our accurate routine is the sum of the last 37 terms of the 42-term Weideman approximation. We refer to our modified Weideman rational approximation as \textit{cpf\_accurate}. Additionally, we changed the order of the Weideman/Humlíček function selector in Schreier's~\cite{Schreier2011} algorithm to achieve further optimization of this method. We refer to our optimized Schreier method as \textit{cpf\_fast}.
\par
Figure~\ref{fig:accuracy} presents the accuracy of both CPF routines. In the considered detuning and broadening parameter range, the \textit{cpf\_accurate} reproduces the exact CPF function with at least seven-digit precision for the real part and ten-digit precision for the imaginary part. The accuracy of the \textit{cpf\_fast} offers only two-digit accuracy in some regions, however, it is still a good, four-digit approximation when the $y$ parameter is greater than $10^{-4}$, which is a typical condition for most applications ($y$ is the imaginary part of the the complex argument, $z=x+i y$, of the CPF). The reference function for generating Fig.~\ref{fig:accuracy} was based on direct integration of Eq.~\eqref{eq:CPF_definifion}.

We implemented both CPF routines in five popular programming languages: Python, Fortran, MATLAB, Mathematica and LabVIEW (the codes were written in the following language versions: Python 3.12.3, Fortran 18, MATLAB 9.13, Mathematica 14 and LabVIEW 2015 SP1). Table~\ref{tab:timings} presents the evaluation time in seconds of both routines on two sample grids: focusing only on the center of the line ($\pm$3FWHM) or also including the far wings ($\pm$25FWHM). As expected, Fortran is the fastest. The numerical tests were performed on a single thread of the 11th Gen Intel Core i9-11900K @ 3.5GHZ CPU with 128 GB RAM (2x 64 GB) available (during the tests, only $\sim$1GB of RAM was occupied).

\subsection{Computer routines for evaluating the mHT profile}
\label{sec:mHTComputerRoutines}

In addition to the CPF routines reported in the previous section, within this work we provide optimized computer routines to evaluate the mHT profile (in the same five programming languages as the CPF: Python, Fortran, MATLAB, Mathematica, and LabVIEW). The present versions of the routines are provided in the Supplementary Materials. Furthermore, the most recent versions are available on the GitHub platform \cite{github_2024}. The routines are supplemented with several simple examples illustrating how to use them. 

The mHT profile routines are based on Eqs.~\eqref{eq:complexqSDHC} and~\eqref{eq:qSDHCLM}. The $\tilde{I}^{*}_{qSDV}(\nu)$ function needed to evaluate the expression from Eq.~\eqref{eq:complexqSDHC} is determined using the CPF, $w(z)$, see Eq.~\eqref{eq:Zform}. The mHT computer code reported here by default uses the \textit{cpf\_accurate} routine for evaluating CPF, see Sec.~\ref{sec:errorFunction} (one can simply modify the mHT code to switch to the \textit{cpf\_fast} routine).  

The $\beta$ correction function (see Sec.~II B) by default is turned off. To activate it one has to set the perturber-to-absorber mass ratio parameter, $\alpha$, to its proper value. The $\beta$ correction is activated when $\alpha<5$, and by default, if not specified explicitly, $\alpha$ is set to 10. Compare the \textit{example\_absorption} and \textit{example\_mHT\_optional\_parameters} code examples in the Supplementary Materials and in Ref.~\cite{github_2024} (the $\beta$ correction is turned off in the first example and turned on in the second one).

\section{The {m}HT profile in the HITRAN database}
\label{sec:HITRANparametersAndHAPI}

In this section, we discuss how the mHT profile is introduced into the upcoming release of the new edition of the HITRAN database~\cite{Gordon2025}. In Section~\ref{sec:tempDep}, we discuss the representation of the temperature dependencies of the mHT profile line-shape parameters. In Section~\ref{sec:HAPI}, we show how the mHT profile is integrated with HAPI and demonstrate an example of using it.

\subsection{Representation of the temperature dependencies of the mHT profile line-shape parameters in the HITRAN database}
\label{sec:tempDep}

The analysis from Ref.~\cite{Stolarczyk2020} concluded that the double-power-law (DPL) representation of temperature dependencies of collision-induced line-shape parameters of the mHT profile gives a better overall approximation of the temperature dependencies than the other approaches. Moreover, the DPL also requires fewer parameters and its structure is much simpler and more self-consistent \cite{Stolarczyk2020}. In this work, we follow the conclusions of Ref.~\cite{Stolarczyk2020} and recommend the DPL parametrization as the most general temperature dependence representation in HITRAN. Note, however, that in many cases (e.g., when the temperature dependencies are available in a narrow temperature range or the temperature dependence is weak) the simpler functions such as the single power law or a linear function suffice and are adopted in HITRAN. In Table~\ref{tab:DPL}, we present the full list of the seven line-shape parameters (the six line-shape parameters for an isolated line, see Eq.~(\ref{eq:parametersA}), plus the line-mixing parameter, $y$) with the names and definitions of their DPL temperature-dependence coefficients that are adopted in HITRAN within its flexible relational structure ~\citep{Hill2013,Hill2016}. In this approach, a set of 28 coefficients (four DPL coefficients per each of the seven line-shape parameters) is used to quantify the collision-induced line-shape effects (as listed in Table~\ref{tab:DPL}) per each perturber species.

\subsection{The mHT profile in the HITRAN Application Programming Interface (HAPI)}
\label{sec:HAPI}

Calculating molecular spectra based on a complex line-shape model (such as the mHT profile) and using the corresponding line-shape parameter datasets (including their temperature dependencies) may be challenging for non-expert database users. To make this structure simpler to access, the HITRAN Application Programming Interface (HAPI)~\cite{Kochanov2016} has been developed and recently the mHT profile was adopted in HITRAN. At this time, the mHT profile has only been parametrized in HITRAN for a small set of data; going forward we hope new accurate spectroscopic measurements will enable us to populate more bands and molecules with these parameters. HAPI allows the HITRAN users to download the necessary parameters and simulate the spectra at arbitrary thermodynamic conditions. In this section, we demonstrate an example of using HAPI to model spectra using the mHT profile.

So far the full mHT data (together with their DPL temperature representation) were introduced to HITRAN for the case of molecular hydrogen (the H$_2$ and HD isotopologues perturbed by He and H$_2$) \cite{Stolarczyk2025}. Here we show an example of HAPI (Python based) code that downloads from HITRAN the necessary data and simulates spectra (absorption coefficient) of He-perturbed H$_2$. 
%The following code illustrate how to download from HITRAN the parameters required to simulate the spectra based on the mHT profile. In this example, the downloaded data are saved to the 'HD' file. \textcolor{red}{Panie Nikodemie, jakie rozszerzenie ma ten plik?} \textcolor{magenta}{potrzebne sa 2 pliki: \.header oraz \.data} The following code illustrates how to simulate the absorption coefficient as a function of wavenumber.
The user needs to create a directory for data storage. In our case it is \textit{hitran\_data}. Subsequently, the user must link it with the database and fetch the desired molecule information in the selected parameterization. Note that \textit{HD} is the name of the local files (2 files are created: HD.header and HD.data), while the HD molecule in the HITRAN database is referred to with its isotopologue ID, ISO\_ID=115.
%\begin{minted}[mathescape]{python}
%from hapi import *
%db_begin ('hitran_data')
%fetch_by_ids('HD',115,
%ParameterGroups=['160-char','mHT'])
%\end{minted}
\begin{verbatim}
from hapi import *
db_begin ('hitran_data')
fetch_by_ids('HD',115,
ParameterGroups=['160-char','mHT'])
\end{verbatim}
After specifying the pressure, temperature as ordinary Python lists \textit{press} and \textit{temp}, and \textit{xmin} and \textit{xmax} for the wavenumber span and \textit{step} for the distances between $x$-axis points, the following code is used to generate the desired spectra with HAPI:
%\begin{minted}[mathescape]{python}
%from hapi import *
%db_begin ('hitran_data')
%nu,coef = absorptionCoefficient_mHT(
%    SourceTables='HD',
%    Diluent={'He':1.0},
%    WavenumberRange=[xmin,xmax],
%    WavenumberStep=step,
%    Environment={'p':press,'T':temp},
%    HITRAN_units=True)
%\end{minted}
\vspace{1em}
\begin{verbatim}
from hapi import *
db_begin ('hitran_data')
nu,coef = absorptionCoefficient_mHT(
    SourceTables='HD',
    Diluent={'He':1.0},
    WavenumberRange=[xmin,xmax],
    WavenumberStep=step,
    Environment={'p':press,'T':temp},
    HITRAN_units=True)
\end{verbatim}
A description of the instructions used can be found in Ref.~\cite{Kochanov2016}. In Figure~\ref{fig:Sbranch}, we show an example of mHT spectra simulated with HAPI, the S branch of the fundamental band in He-perturbed HD at three different temperatures ($T=1000$, $300$, and $50$~K) at $p=0.2$~atm. Recently, the use of HAPI with the mHT profile was demonstrated in another example of HD lines perturbed by an H$_2$ and He mixture bath \cite{Jozwiak2024}.

%The parameters of the new profile were incorporated into the relational structure of the HITRAN database, and  are available through HITRAN\textit{online} (\url{https://hitran.org}) and .
%To access these parameters, HITRAN users can proceed to HITRAN\textit{online} to create a customized output file when downloading line-by-line data. 

\section{Conclusion}
\label{sec:conclusion}

In this work, we analyzed some problematic features of the Hartmann-Tran (HT) line-shape.
The two most important ones are the singular behavior of the temperature dependencies of the velocity-changing parameters when the shift parameter crosses zero and the difficulty in evaluating the velocity-changing parameters for mixtures. We demonstrated a straightforward way to eliminate the above-mentioned problem. We refer to such a refined profile as the modified Hartmann-Tran (mHT) profile. The computational cost of evaluating it is the same as for the HT profile. We gave a detailed description of the mHT profile (also including line mixing) and discussed the representation of its parameters, together with their DPL temperature parametrization. We presented an efficient algorithm for evaluating the mHT profile and provided corresponding computer codes in several programming languages: Fortran, Python, MATLAB, Mathematica, and LabVIEW. We recommend using the mHT profile with the DPL temperature dependencies in the HITRAN database. The current HT parametrization will be retained for consistency reasons. At the moment only the HITRAN data for molecular hydrogen can be updated in the full representation of this new parametrization. The other molecules that have HT data do not contain the correlation parameters (with some exceptions) and do not have sufficient data to allow for a DPL representation. The parameters will be duplicated into the mHT parametrization. Going forward, we encourage scientists to fit their data using the mHT profile rather than with the HT profile. We also discussed the corresponding update of the HITRAN Application Programming Interface (HAPI).

\section*{Acknowledgments}

The project was supported by the National Science Centre in Poland through Project Nos. 2022/46/E/ST2/00282 (P.W, M.S.), 2019/35/B/ST2/01118 (H.J, N.S), 2023/51/B/ST2/00427 (D.L.), 2021/41/B/ST2/00681 (R.C.), 2020/39/B/ST2/00719 (Ag.C.). For the purpose of Open Access, the authors have applied a CC-BY public copyright licence to any Author Accepted Manuscript (AAM) version arising from this submission. NIST acknowledges funding from the National Aeronautics and Space Administration (NASA) [contract NNH20ZDA0001N-OCOT]. An.C. and L.G. acknowledge the EURAMET Metrology Partnership project “PriSpecTemp” [grant numbers 22IEM03]. This project has received funding from Metrology Partnership program co-financed by the Participating States and the European Union’s Horizon 2020 research and motivation programme. We gratefully acknowledge Polish high-performance computing infrastructure PLGrid (HPC Center: ACK Cyfronet AGH) for providing computer facilities and support within computational grant no. PLG/2024/017376. Created using resources provided by Wroclaw Centre for Networking and Supercomputing (\url{http://wcss.pl)}.

\appendix

\section{The complex Dicke parameter for mixtures}
\label{sec:app1}

In general, to handle the velocity-changing collisions in a line-shape model for a mixture of perturbers, one must calculate the velocity-changing operator as a weighted sum of the velocity-changing operators for all perturbing species, see Sec.~V.B in Ref.~\cite{Wcislo2014} (the full approach was implemented in a few recent works \cite{Wcislo2015PRA2, Slowinski2020, Stolarczyk2023}). However, when the same velocity-changing operator is used for all the perturbing species (which is the case for both HT and mHT profiles), the problem simplifies to calculating a weighted sum of the complex Dicke narrowing parameters, $\tilde{\nu}_{opt,i}$,
\begin{equation}
    \nu_{opt}=\sum\limits_{i}p_i\tilde{\nu}_{opt,i},
    \label{eq:app1}
\end{equation}
where $p_i$ is the partial pressure of the $i$-th constituent. In the case of the HT profile, $\nu_{opt}$ and $\tilde{\nu}_{opt,i}$ depend on the speed of the active molecule. 

\subsection{mHT profile}

In the case of the mHT profile, the problem of calculating $\nu_{opt}$ for mixtures is simple and straightforward. $\nu_{opt}$ is a direct sum of its real and imaginary parts, $\nu_{opt}=\nu_{opt}^r+i\nu_{opt}^i$, hence it follows from Eq.~(\ref{eq:app1}) that $\nu_{opt}^r$ and $\nu_{opt}^i$ for mixtures are calculated as simple weighted sums (exactly the same as for the four other line-shape parameters from list~(\ref{eq:parameters})) 
\begin{equation}
    \nu_{opt}^r=\sum\limits_{i}p_i\tilde{\nu}_{opt,i}^r,
    \label{eq:app2}
\end{equation}
\begin{equation}
    \nu_{opt}^i=\sum\limits_{i}p_i\tilde{\nu}_{opt,i}^i.
    \label{eq:app3}
\end{equation}

\subsection{HT profile}

In the case of the HT profile, the problem of calculating $\nu_{opt}$ for mixtures is much more complicated. For the $i$-th mixture constituent, the complex Dicke narrowing parameter is parameterized as
\begin{equation}
\begin{split}
\tilde{\nu}_{opt,i} & = \tilde{\nu}_{VC,i}-\eta_i(\gamma_{0,i}+i\delta_{0,i})+\\
& -\eta_i(\gamma_{2,i}+i\delta_{2,i})(v^2/v_m^2-3/2).
    \label{eq:app4a}
\end{split}
\end{equation}

On the one hand, substituting Eq.~(\ref{eq:app4a}) into Eq.~(\ref{eq:app1}), we get
\begin{equation}
\begin{split}
\nu_{opt} & = \sum\limits_i p_i\tilde{\nu}_{VC,i}-\sum\limits_i p_i\eta_i(\gamma_{0,i}+i\delta_{0,i})+\\
& -(v^2/v_m^2-3/2)\sum\limits_i p_i\eta_i(\gamma_{2,i}+i\delta_{2,i}).
    \label{eq:app4}
\end{split}
\end{equation}
On the other hand, $\nu_{opt}$ can be written as in Eq.~(\ref{eq:HTPnuOpt}), but the parameters describe the effective values for a mixture
\begin{equation}
\nu_{opt}=\nu_{VC}-\eta(\Gamma_0+i\Delta_0)-\eta(\Gamma_2+i\Delta_2)(v^2/v_m^2-3/2).
    \label{eq:app5}
\end{equation}
A direct comparison of speed-independent and speed-dependent terms in Eqs.~(\ref{eq:app4}) and (\ref{eq:app5}) gives
\begin{equation}
\begin{split}
&\nu_{VC}-\eta(\Gamma_0+i\Delta_0)  =\\
&\sum\limits_i p_i\tilde{\nu}_{VC,i}-\sum\limits_i p_i\eta_i(\gamma_{0,i}+i\delta_{0,i}) ,
    \label{eq:app6}
\end{split}
\end{equation}
\begin{equation}
\begin{split}
\eta(\Gamma_2+i\Delta_2) & =\sum\limits_i p_i\eta_i(\gamma_{2,i}+i\delta_{2,i}).
    \label{eq:app7}
\end{split}
\end{equation}
Equation~(\ref{eq:app7}) directly gives the expression for the $\eta$ parameter given in Eq.~(\ref{eq:etaSum}). Equation~(\ref{eq:app6}) with $\eta$ substituted with the expression from Eq.~(\ref{eq:etaSum}) gives $\nu_{VC}$, see Eq.~(\ref{eq:nuVCSum}).

\section{Computer routines for efficient evaluation of the mHT profile}
\label{sec:app2}
This appendix summarizes the details of the \textit{cpf\_accurate} and \textit{cpf\_fast} routines.  
 \par
 \subsection{\textit{cpf\_accurate}}
This algorithm is based on the Weideman's 64-term  rational approximation~\cite{Weideman1994}, which we optimized specifically for the purpose of generating the mHT function. We reduced the approximation terms to 42 and pre-calculated the $L$ constant ($L=\sqrt{N/\sqrt{2}}$, where $N$ is the number of approximation terms, here $N=42$). Because $N$ was fixed, we also pre-calculated the coefficients of the $a$ table, which are usually obtained through FFT. The elements of the $a$ table are the coefficients of the polynomial $p$. We found that the first five elements of $a$ are below $10^{-16}$ and we truncated them, which is identical to substituting these low values with zeros. We found that this operation has no effect on the result of the cpf function and it accelerates the calculation by reducing the number of operations.
The construction of the polynomial $p$ was usually done with \textit{for} loops. In our codes we applied Horner's scheme to accelerate the calculation of $p$. Ultimately, for the evaluation of the cpf value from the $L$, $p$, and the input parameters, we rearranged the operations and reduced one division for further acceleration. We also pre-calculated the inverse square root of $\pi$ to avoid evaluating it with every run of the code. 

 \subsection{\textit{cpf\_fast}}
This algorithm is based on the approach of dividing the complex plane of the CPF input parameters into subregions, an idea initially proposed by Humlíček~\cite{Humlicek1982}. Although Humlíček divided it into four subregions, suggesting different approaches for each of them, Schreier developed another approach~\cite{Schreier2011}, which is often referred to as \textit{hum1\_wei24}. He used Humlíček's approximation in its first subregion ($|x|+y<15$, where $x$ and $y$ are the CPF real and imaginary inputs, respectively), and applied the 24-term Weideman approximation~\cite{Weideman1994} in the remaining three ranges.
\par
Within this work, we optimized the Weideman approximation part the same way as we did for the \textit{cpf\_accurate}, i.e. we pre-calculated the $L$ and $a$ constants with $N=24$ this time, as well as $\sqrt{\pi^{-1}}$, and we rearranged the operations in the final line of the Weideman part of the code. We also simplified the selection of the methods dependent of the region. Because the Humlíček part of the code is faster, the function will not load the Weideman part unless it is required.  

%\subsection{\textit{mHT}}
%The computer codes for calculating the \textit{mHT} function we present in this article are inspired by the Fortran subroutines from Ref~\cite{Tran2013}, which generate the partially-Correlated quadratic-Speed-Dependent Hard-Collision (pCqSDHC) profile in the Hartmann-Tran parameterization. This code has later been adopted to HAPI library and rewritten in python~\cite{Kochanov2016}.
%\par
%Besides switching from the $\eta$ and $\nu_{vc}$ to the $\widetilde{\nu}_{\rm opt}^{\rm r}$ and $\widetilde{\nu}_{\rm opt}^{\rm i}$ parameterization, our code differs significantly from the one present in the HAPI library. Most importantly, we added the beta correction function to the code, 

%\begin{itemize}
%    \item cleared the code in general
%    \item we optimized the order of ifs
%    \item resigned from library uses wherever applicable
%    \item gave actual values of mathematical parameters like pi
%    \item pre-calculated variables which are used several times
%    \item where applicable, compiled the code on the processor in interpreted languages
%\end{itemize}

\section{Refinements to the DPL fitting procedure}
\label{sec:app3}
The double-power-law (DPL) temperature dependency representation has been adopted in the HITRAN database since 2020~\cite{Stolarczyk2020}, in the following form:
\begin{multline}
    {\rm Param.}(\rm T)= \\Coef.1({\rm T_{\rm ref}}/\rm T)^{Exp.1}+Coef.2({\rm T_{\rm ref}}/\rm T)^{Exp.2}.
\end{multline}
We have identified several issues regarding the practical applications of the DPL function. 
\par
Since the two terms of the DPL function are mathematically identical, the coefficients were not uniquely identified, which was an issue from the perspective of database storage.
\par
Another issue was the unbounded nature of the DPL fitting procedure. The fitted coefficients would sometimes evaluate to huge, similar values with opposite signs. The lack of bounds led to extremely long fitting times as numerous local minima of the fitting function corresponded to points where the two terms compensated in different configurations. Not only did it lead to unequivocal, unstable solutions to the same fitting problems but also the DPL coefficients were several orders of magnitude greater than the original line-shape parameters. Because of their similarity and magnitude, storing the coefficients with sufficient numerical precision to accurately represent the temperature dependencies required more database space. In some cases, 16 digits were required to reconstruct the temperature dependencies.
\par
To solve these issues and guarantee the homogeneity of the database entries we set the following conditions for the definition and fitting procedure of the DPL function:
\begin{itemize}
    \item $Coef.1>Coef.2$, which guarantees unique definition of the DPL coefficients.
    \item $\frac{|Coef.2|-|Coef.1|}{|Coef.1|}>10^{-3}$. This constraint ensures that Coef.1 and Coef.2 are sufficiently different in absolute value. Specifically, their difference must be large enough to affect at least the fourth significant digit. This bound prevents the method from assigning nearly identical large values with opposite signs, improving computational efficiency and reducing storage requirements.
    \item The HITRAN database will provide DPL coefficients with up to eight significant figures.
\end{itemize}

%\section*{References}
%\bibliographystyle{apsrev4-1}
\bibliographystyle{jqsrt_doi_nourl}
\bibliography{HITRANlineshape.bib}

\end{document}